\documentclass[12pt]{iopart}

\usepackage{iopams}
\usepackage{setstack}
\usepackage[english]{babel}
\usepackage{cite}
\usepackage[pdftex]{graphicx}
\usepackage[pdftex]{hyperref}
\usepackage{lscape} %Provê o ambiente landscape e tudo dentro dele ficará na vertical
\usepackage{pdflscape} 
\usepackage{rotating}
\usepackage[usenames]{color}
\usepackage[normalem]{ulem}

\begin{document}

\title[Information and quantum theories: an analysis in one-dimensional systems]{Information and quantum theories: an analysis in one-dimensional systems}

\author{Wallas S. Nascimento, Marcos M. de Almeida and Frederico V. Prudente }

\address{Instituto de F\'{\i}sica, Universidade Federal da Bahia, 40170-115, Salvador, Bahia, Brazil.}
\ead{wallassantos@gmail.com, marcosma@ufba.br, prudente@ufba.br}
\vspace{10pt}

% \begin{indented}
% \item[]Date:
% \end{indented}

\begin{abstract}
We pedagogically present the information theory as originally established, explaining its essential ideas and paying attention to the expression employed to measure the amount of information. Also we discussed relationships between information and quantum theories. In this context we determined the information entropies on position, $S_x$, and momentum, $S_p$, spaces, besides the entropy sum $S_t$. Here, we use the modified entropic expressions that are dimensionally consistent. We provide original explanations for the behaviors of the $S_x$, $S_p$ and $S_t$ values analyzing the probability densities and by means of the normalization constants and properties of Fourier transform. We validate the entropic uncertainty relation and inspect the standard deviation and information entropies as measures of quantum uncertainty. The systems of interest unidimensional one-particle quantum systems in ground and excited states. 
\end{abstract}

\vspace{2pc}

\submitto{Eur. J. Phys. https://doi.org/10.1088/1361-6404/ab5f7d}
%\maketitle

\section{Introduction}

The understanding and interpretation of the quantity known as entropy include at least three areas of knowledge, namely: thermodynamics, statistical mechanics and information theory. The entropy arises in the thermodynamics' scope~\cite{greiner,caminhosparaentropia}, but with the atomistic assumption and statistical methods such concept gains a new meaning~\cite{pathria,diu}. It also emerges in the communications through use of the information entropy or Shannon entropy~\cite{livro_teoria_matemcatica_da_comunicacao_original,rioul}. Information entropies on position, $S_r$, and momentum, $S_p$, spaces, besides the entropy sum $S_t$, connect information theory and quantum mechanics~\cite{livro_sen_shannon,artigoreferenciashannon}. In this way, concepts and ideas find applicability in different areas aside those originally defined.

Initially identified as an autonomous area, the mathematical theory of communication or information theory now has its fundamental concepts utilized in distinct fields of expertise, mainly through the informational entropy~\cite{conceitodeinformacao,ganhodeinformacao,machta}. The information theory in the atomic, molecular and chemical physics context has generated a reasonable number of analyses~\cite{robinquantumwell,lagunasagar,entropyhyperbolicalpotential,entropyrosenmorsepotential,rojasaquino,sarsa,davidcenteno,shannongaiola}. Works about ions in plasma environments~\cite{plasmashannon} and correlation measurements~\cite{shannon_correlacao2, shannon_correlacao} show great results in favor of the informational treatment. The examination of strong confinement regime~\cite{wallas_fred_internacional} and systems constrained by a dielectric continuum~\cite{confineddie} apply the informational language. Also, the entropic expressions analyze the phenomena of localization or delocalization of the probability densities~\cite{aquino, locadeloca}. 

One-dimensional quantum systems are a standard topic of the quantum theory studied in undergraduate courses of physics and related areas. They present important features as benchmark systems: have exact solutions, reveal the existence of non-classical effects and do not provide many difficulties in solution as do higher dimension systems. The research for the one-dimensional systems using the informational entropy presents contributions to infinite potential well (particle in a box)~\cite{relacoessomaentropica2, 
particula_caixa_shannon2} and to harmonic potential (harmonic oscillator)~\cite{wallas_fred_educacional,
uncertainties_of_the_confined_harmonic}.

The goal of this work is to pedagogically show for teachers and students relationships between information and quantum theories. We consider the connection provided by probability densities, emphasizing the modified entropic expressions that are dimensionally consistent. Applications to one-dimensional quantum systems are investigated.

The paper is organized as follows. In section~\ref{Information Theory} we present the information theory as originally established, explaining its essential ideas and paying attention  to the expression employed to measure the amount of information. In section~\ref{articulation} we identify connections between information and quantum theories, highlighting certain subtleties. In section~\ref{interest systems} we discuss the unidimensional one-particle quantum systems of interest, while in Section~\ref{Results} we present and analyze our results that are compared, when available, with those previously published. Finally, in section~\ref{conclusions} we summarize the main aspects of the current study. In appendix, we perform a dimensional analysis of the entropic expressions. In the appendix A and B, we discussed how to get $S_t$ expression and we perform a dimensional analysis of the entropic expressions, respectively. All formalisms presented in this paper will be developed for one-dimensional systems.

\section{Information theory} {\label{Information Theory}}

A mathematical theory of communication or information theory arose with report of Claude Shannon in 1948~\cite{shannonoriginal}. Other works are significant for understanding the problem: Harry Nyquist's considerations~\cite{nyquist}, which suggests a quantity of telegraphic data, and Ralph Hartley~\cite{hartley}, that delimits the meaning of information and shows its measurement by a logarithmic function. Warren Weaver's explanation expands the applications of Shannon researches to include a spectrum of processes such as oral transmissions, music and photography~\cite{livro_teoria_matemcatica_da_comunicacao_original}.

The Communication is considered as a process in which one mechanism affects the other through a message. In this background, information is a measure of the choice of a message within an available repertoire. The essential question of the information theory is how to replicate at a destination point a message (or as similar as possible) transmitted from a point of origin~\cite{livro_teoria_matemcatica_da_comunicacao_original}.

\begin{figure}[h]
\begin{center}
\scalebox{0.60}{ \includegraphics[trim=2cm 0cm 0cm 0cm]
{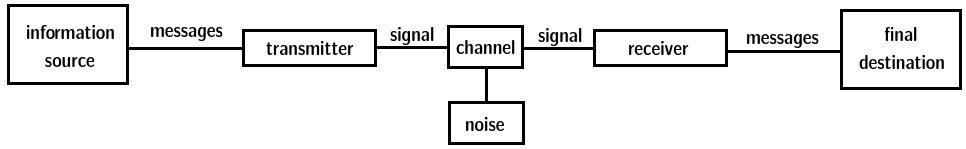} }
\end{center}
\caption{Diagram of a general communication system.}
\label{sistemacomunicao}
\end{figure}

The model that defines a general communication system is illustrated in figure~\ref{sistemacomunicao}. In this diagram, the \textbf{information source} chooses a message from a possible group of them, so the \textbf{transmitter} codes the message into a signal that is sent by the \textbf{communication channel}. The message in communication channel may be influenced by the \textbf{noise}, characterized as external changes imposed on the signal. The \textbf{receiver} decodes the original message to be delivered in \textbf{final destination}.

In sending and receiving process of a message, the semantic aspects are secondary ones. The expression employed to measure the amount of information generated in a message by a discrete information source is~\cite{shannonoriginal},
\begin{equation}
S\left(\{p_i\}\right)=-{\sum_{i=1}^{ j }} {p_i}\log_2 (p_i) \ ,
\label{shannondis}
\end{equation} 
where $\{p_i\}$ represents the group of complete messages, $j$ is the number of messages and $p_i$ is the probability of occurrence of $i$-th message. These probabilities are constrained by normalization condition, $\sum\limits_{i=1}^{{ j }} {p_i}=1$, and non-negative condition, $p_i \geq 0$, $\forall \, i$. To $i=a$ and $p_a=0$, implies $p_a\log p_a \equiv  0$. In case of a continuous information source \cite{shannonoriginal}, equation~(\ref{shannondis}) is written as 
\begin{equation}
S(p(\alpha))=  - \int\limits_{-\infty}^{\infty} d\alpha \ p(\alpha) \log_2 p(\alpha)  \ , 
\label{shannoncontinuo}
\end{equation} 
where $p(\alpha)$ is a probability density in function of continuous $\alpha$ variable, also constrained by normalization condition, 
$\int\limits_{-\infty}^{\infty} d\alpha \ p(\alpha) =1$, and non-negative condition, $p(\alpha) \geq 0$, $\forall \, \alpha$. The values provided by equation~(\ref{shannoncontinuo}) may be negative (see pg. 631 in reference~\cite{shannonoriginal}), where $p(\alpha)$ is bigger than 1 in some region of the $\alpha$-domain. The expressions~(\ref{shannondis})~and~(\ref{shannoncontinuo}) are known as information entropy or Shannon entropy.\footnote{The relationship between Boltzmann and Shannon entropies is a controversial point and similarities are not so clear. Although definitions of both quantities are based on probability distributions, anyone should take care comparing them. In this way, there are peculiar forms of comparing these distinct quantities~\cite{notions_of_entropy, thermodynamicbasedonshannonentropy}.}

The logarithmic base in equations~(\ref{shannondis})~and~(\ref{shannoncontinuo}) specifies the informational unit, \textit{e.g.}, hartleys and nats for 10 and neperian number logarithmic bases, respectively. When communication processes adopt base 2, informational unit is the bit (binary digit). Note that a base change involves only a variation of scale.

\section{Connection between information and quantum theories} {\label{articulation}}

Max Born, in the probabilistic interpretation of quantum mechanics~\cite{bornoriginal}, proposed the quantity $\rho(x)$ to be related to the probability density on position space and assigned it, in terms of Schr\"odinger equation solution $\psi (x)$, as $\rho(x) = \arrowvert \psi (x)\arrowvert^2$. The Fourier Transform of the function $\psi (x)$ is defined by ${\widetilde{\psi}(p)}$. In the same way, we establish a probability density on momentum space as $\gamma(p)={\arrowvert \widetilde{\psi}(p) \arrowvert}^2$. In Quantum Theory, $\rho(x)$ has dimension of inverse of length and $\gamma(p)$ has a dimension of inverse of momentum. 

The dimensional adequate information entropies on position, $S_x$, and on momentum, $S_p$, spaces are given by~\cite{wallas_fred_internacional}:
\begin{equation}
S_x = -\int  dx \ \rho(x) \ln \left( {a_0} \ \rho(x) \right)
\label{entropia_posicao_x}
\end{equation}
and
\begin{equation}
S_p = -\int  dp \  \gamma(p)\ln \left( {\left( \frac{\hbar}{a_0} \right)} \ \gamma(p)  \right) \ .
\label{entropia_momento_x}
\end{equation}
Here, the probability densities $\rho(x)$ and $\gamma(p)$ are normalized to unity. The equations~(\ref{entropia_posicao_x})~and~(\ref{entropia_momento_x}) are characterized with the fundamental physical constants $a_0$, Bohr radius, and $\hbar$, reduced Planck constant.  

In the framework of union between information theory and quantum mechanics we obtain the entropic uncertainty relation. This relation is derived from entropy sum of $S_x + S_p$ of the non-commuting position $\mathcal{X}$ and momentum $\mathcal{P}$ observables~\cite{relacaodeincertezainformacao}. So, we have~\cite{wallas_fred_internacional}
\begin{equation}
S_t=S_x+S_p= - \int \int dx \ dp \  \rho(x)  \ \gamma(p) \ \ln( \ \hbar \ \rho(x) \ \gamma(p) \ ) \ \geq (1+\ln\pi) \ .
\label{Stx}
\end{equation}
Note that $S_t$ value is bounded because displays a minimum value for entropy sum. In the Appendix A we discussed how to get $S_t$ expression.

An adequate dimensional analysis in  expressions~(\ref{entropia_posicao_x}),~(\ref{entropia_momento_x})~and~(\ref{Stx}) are guaranteed by the fundamental physical constants $a_0$ and $\hbar$. We examine in the Appendix the dimensional balance of the entropic expressions in this work. We show that modified information entropies $S_x$ and $S_p$, besides of modified entropic uncertainty relation are dimensionally adequate. The dimensional analysis does not consider any possible unit, thus the expressions proposed in reference~\cite{wallas_fred_internacional} have a more general aspect than the relation regularly employed for the information entropies, \textit{e.g.}, reference~\cite{livro_sen_shannon}. Using atomic units in the modified relationship recovers its conventional way, but, now with a dimensionally convenient expression.

In quantum theory, it is known that the measure of any two non commuting observables $\mathcal{A}$ and $\mathcal{B}$ can only be done within a limit of accuracy. These quantitative descriptions are recognized as uncertainty relations. In this sense, the Heisenberg's uncertainty principle presents this unpredictability in a conceptually way~\cite{heisenberg,gillespie}. In particular, for position $\mathcal{X}$ and momentum $\mathcal{P}$ observables, with their respective quantum operators $\hat{X}$ and $\hat{P}$, Kennard rigorously establishes the relation~\cite{kennard,watanabe}
\begin{equation}
\Delta \hat{X} \Delta \hat{P} \geq \frac{\hbar}{2} \ .
\label{princípiodeh}
\end{equation}
The standard deviations are
\begin{equation}
\Delta \hat{X} = \sqrt{ {\langle x^2 \rangle} - {\langle x \rangle}^2} \ \ {\rm and} \ \ \Delta \hat{P} = \sqrt{ {\langle p^2 \rangle} - {\langle p \rangle}^2} \ , 
\end{equation}
where 
\begin{equation}
{\langle x \rangle} =\int \ dx \ x \rho (x) \ \ , \ \ { \langle x^2 \rangle} =\int \ dx \ x^2 \rho (x)
\end{equation}
and
\begin{equation}
{\langle p \rangle} =\int \ dp \ p \gamma(p) \ \ , \ \ { \langle p^2 \rangle} =\int \ dp \ p^2 \gamma(p) \ . 
\end{equation}
The Kennard's relation~(\ref{princípiodeh}) establishes that it is not possible to simultaneously measure position and momentum of a particle with arbitrary precision.

$\Delta \hat{X}$ or $\Delta \hat{P}$ quantities are dispersion (or spread) measures of the probability distributions in relation to the mean value ${\langle x \rangle}$ or ${\langle p \rangle}$. $S_x$ and $S_p$ entropies are also a type of dispersion or spread measures. However, in this specific case,  they are calculated without taking into account reference points in the probability distributions, being therefore a genuine spread measure of the probability distributions. In this sense, the standard deviation and information entropies are measure of uncertainty (localization or delocalization of the particle in the space) with different characteristics. The uncertainties can also be quantified by means of the Fisher information~\cite{WU2019126033} and Tsallis and R\'enyi entropies~\cite{PhysRevA.84.034101}, among others. The topic of which quantities are most consistent for to measure quantum uncertainty is object of discussion in the literature~\cite{discussaoincerteza1, discussaoincerteza2, revisaorelacaoincertezaentropica,incertezaentropiamarj}.

Uncertainty relations~(\ref{Stx})~and~(\ref{princípiodeh}) reach their minimum values with adoption of Gaussian-type wave functions such as in the ground state of harmonic oscillator~\cite{sakurai,wallas_fred_educacional}. The entropic uncertainty relation is considered as a stronger version of the Kennard's relation, in the sense that from relation~(\ref{Stx}) we can deduce the relation~(\ref{princípiodeh})~\cite{relacaodeincertezainformacao}. Still, uncertainty relations in terms of $S_x$ and $S_p$ are proposed to deal with situations where the Heisenberg's uncertainty principle or Kennard's relation presents sensitivities~\cite{relacoessomaentropica2, particula_caixa_shannon2}, as for instance, in the study of separable phase-space distribution~\cite{uncertainties_of_the_confined_harmonic, shannon_wigner}.

\section{Systems of Interest} {\label{interest systems}}
  
The time-independent Schr\"odinger equation is written as 
\begin{equation}
 -\frac{\hbar^2}{2m} \frac{d^{2}\psi(x)}{d{x}^{2}} + V(x)\psi(x)=E\psi(x) \ ,
\label{schodingerumadimensao}
\end{equation}
where $m$ is the mass of the particle, $E$ is the energy of the stationary state and $V(x)$ is the potential function. 

A complete statement of the question is set when establishing the potential function $V(x)$ in equation~(\ref{schodingerumadimensao}) and boundary conditions for wave function. The potential functions of interest in this paper are specified in the subsections~\ref{Harmonic potential}~and~\ref{infinite potential well}.

\subsection{Harmonic potential}{\label{Harmonic potential}}

As an initial approximation, this model expresses the relative motion of atoms in molecules and solids. The harmonic potential is given by
\begin{equation}
V(x)=\frac{1}{2}m\omega^2x^2 \ ,
\label{potencialharmonico}
\end{equation}
where $\omega$ is the angular frequency of the classical oscillator and $x$ is the displacement of the mass $m$ regarding the equilibrium position in origin of coordinate framework. From now we call this system a harmonic oscillator. The angular oscillation frequency $\omega$ relates to force constant $k$ by the expression $\omega=\sqrt{k/m}$. 

The resolution of equation~(\ref{schodingerumadimensao}) for the potential function~(\ref{potencialharmonico}) can be done by different procedures such as the algebraic~\cite{sakurai} and the analytical~\cite{griffithsingles} ones. The eigenvalues and eigenfunctions are respectively 
\begin{eqnarray}
E_n&=&\hbar \omega (n+1/2) \\
\psi_n(x)&=&A_n e^{-\frac{\beta x^2}{2}}H_n(\sqrt{\beta } x) \ ,
\label{solucaogeralhermite}
\end{eqnarray}
where $A_n=2^{-n/2}\pi^{-1/4}(n!)^{-1/2}\beta^{1/4}$ is the normalization constant, $\beta$ parameter is $m\omega/\hbar$ and $H_n(\sqrt{\beta }x)$ represents the Hermite polynomials~\cite{arfken}. The quantum number $n$ takes non-negative values and specifies the quantum state of the system.

\subsection{Infinite potential well} \label{infinite potential well}

The infinite potential well is defined by
\begin{equation}
V(x)=  \left \{ \begin{array}{ccl}
\infty & & {\rm to} \ \ |x| \ge x_c/2  \\
 0 & & {\rm to} \ \ |x| < x_c/2  
\end{array} \right. ,
\label{potencial_caixa}
\end{equation}
where $x_c$ is the confinement distance (width of the box). From now we call this system a particle in a box. For the $x$ range of values between $-x_c/2$ and $x_c/2$, the particle is free. Boundary conditions, $\psi( x= \pm x_c/2) =0$, forces the confinement. 

The general solution of equation~(\ref{schodingerumadimensao}) for potential function~(\ref{potencial_caixa}) is given by~\cite{eisberg},
\begin{equation}
\psi(x)=Ae^{ikx}+Be^{-ikx} \ .
\label{solucaogeral}
\end{equation} 
By imposing the boundary conditions in equation~(\ref{solucaogeral}) and choosing a nontrivial solution, we found  
\begin{equation}
  \psi_n(x)=A_n\cos(k_nx)
\label{solucao_cos}
\end{equation} 
and  
\begin{equation}
  \psi_n(x)=B_n \sin(k_nx),
\label{solucao_sen}
\end{equation} 
with the eigenvalues given as
\begin{equation}
E_n=\frac{\pi^2\hbar^2 n^2}{2mx_c^2}.    
\end{equation}
Normalization constants $A_n$ and $B_n$ are equal and independent of the state, they only depend on $x_c^{-1/2}$. The parameter $k_n=n\pi/(2x_c)$ is identified as the wave number. Furthermore, $n$ specifies the quantum number and determines the fundamental and excited states of the system. The cosine type solution adopts $n=1,3,5,\ldots$, while for the sine type, $n$ takes the values $2,4,6,\ldots$.

\section{Results and discussion}{\label{Results}}

In subsections~\ref{Harmonic oscillator result}~and~\ref{Infinite potential well result}, we examine the data for the harmonic potential and for the infinite potential well. We use the atomic units (a.u.) system, a usual one in atomic and molecular physics works. This system uses the mass, $m_e$, and the elementary charge, $e$, of the electron, the constant of electrostatic force $1/4 \pi \varepsilon_0$ and the reduced Planck constant $\hbar$ as standard units of their quantities. Atomic unit system, beyond simplifying main equations in quantum theory for atoms and molecules, has computational advantages in numerical computations. We used the software Maple13 to perform the calculations. In all calculations we consider $m=m_e$. 

\subsection{Harmonic potential}{\label{Harmonic oscillator result}}

We investigated the three lowest energy states $n=0, 1$ and $2$ of harmonic oscillator. The values of modified information entropies $S_x$ e $S_p$ as a function of $\omega$ are presented in table~\ref{energias_entropias_frequencias}, jointly with the entropy sum $S_t$. With a decrease of $\omega$ value, $S_x$ value increases, on the other hand, $S_p$ value decreases. Moreover, $S_t$ is constant with $\omega$ and increases with $n$. The reference~\cite{gardre1985} establishes the values of $S_x$ and $S_p$ for the first six quantum states of the harmonic oscillator for $\omega$~=~1.0000 a.u.. Such study indicates that the information entropies values increase with $n$. The present paper endorses the reference~\cite{gardre1985} and generalizes this characteristic for some values of $\omega$.

%\begin{landscape}
\begin{table}[!ht]
%\footnotesize
\caption{Modified information entropies $S_x$ and $S_p$, besides of the entropy sum $S_t$ as a function of $\omega$ for the harmonic oscillator to three lowest energy states. All values in atomic units system.} 
\label{energias_entropias_frequencias}
\centering
\begin{tabular}{ccccccccccccccccc}
\\ \hline \hline 
 & \vline &  & $S_x$	& & \vline	&		&$S_p$		&	& \vline	&	&$S_t$	&
\\ \hline
$\omega$ & \vline & $n$~=~0 & $n$~=~1	& $n$~=~2& \vline	&$n$~=~0		&$n$~=~1		& $n$~=~2	& \vline	&$n$~=~0	&$n$~=~1	&$n$~=~2
\\ \hline
0.0600& \vline &	2.4791&	2.7494&	2.9053& \vline &	-0.3343	&-0.0640	&0.0919& \vline &	2.1447	&2.6855	&2.9972
\\ \hline
0.0800& \vline &	2.3352&	2.6056&	2.7615& \vline &	-0.1905	&0.0799	&0.2357& \vline &	2.1447	&2.6855	&2.9972
\\ \hline
0.2000& \vline &	1.8771&	2.1474&	2.3033& \vline &	0.2676	&0.5380	&0.6939& \vline &	2.1447	&2.6855	&2.9972
\\ \hline
0.4000& \vline &	1.5305&	1.8009&	1.9568& \vline &	0.6142	&0.8846	&1.0405& \vline &	2.1447	&2.6855	&2.9972
\\ \hline
0.5000& \vline &	1.4189&	1.6893&	1.8452& \vline &	0.7258	&0.9962	&1.1520& \vline &	2.1447	&2.6855	&2.9972
\\ \hline
1.0000& \vline &	1.0724&	1.3427&	1.4986& \vline &	1.0724	&1.3427	&1.4986& \vline &	2.1447	&2.6855	&2.9972
\\ \hline
2.0000& \vline &	0.7258&	0.9962&	1.1520& \vline &	1.4189	&1.6893	&1.8452& \vline &	2.1447	&2.6855	&2.9972
\\ \hline
3.0000& \vline &	0.5231&	0.7934&	0.9493& \vline &	1.6217	&1.8920	&2.0479& \vline &	2.1447	&2.6855	&2.9972
\\ \hline
4.0000& \vline &	0.3792&	0.6496&	0.8055& \vline &	1.7655	&2.0359	&2.1918& \vline &	2.1447	&2.6855	&2.9972
\\ \hline
5.0000& \vline &	0.2676&	0.5380&	0.6939& \vline &	1.8771	&2.1474	&2.3033& \vline &	2.1447	&2.6855	&2.9972
\\ \hline
6.0000& \vline &	0.1765&	0.4468&	0.6027& \vline &	1.9682	&2.2386	&2.3945& \vline &	2.1447	&2.6855	&2.9972
\\ \hline
7.0000& \vline &	0.0994&	0.3698&	0.5257& \vline &	2.0453	&2.3157	&2.4716& \vline &	2.1447	&2.6855	&2.9972
\\ \hline
8.0050& \vline &	0.0323&	0.3027&	0.4586& \vline &	2.1124	&2.3828	&2.5386& \vline &	2.1447	&2.6855	&2.9972
\\ \hline \hline 
%\footnotemark[1]{\footnotesize{$x_c=3,3$ u.a.}}
\end{tabular}
\end{table}
%\end{landscape}

\begin{figure}[b] % aqui começa o ambiente figura
\centering % este comando é para centralizar a figura
\includegraphics[width=9cm, height=7cm]{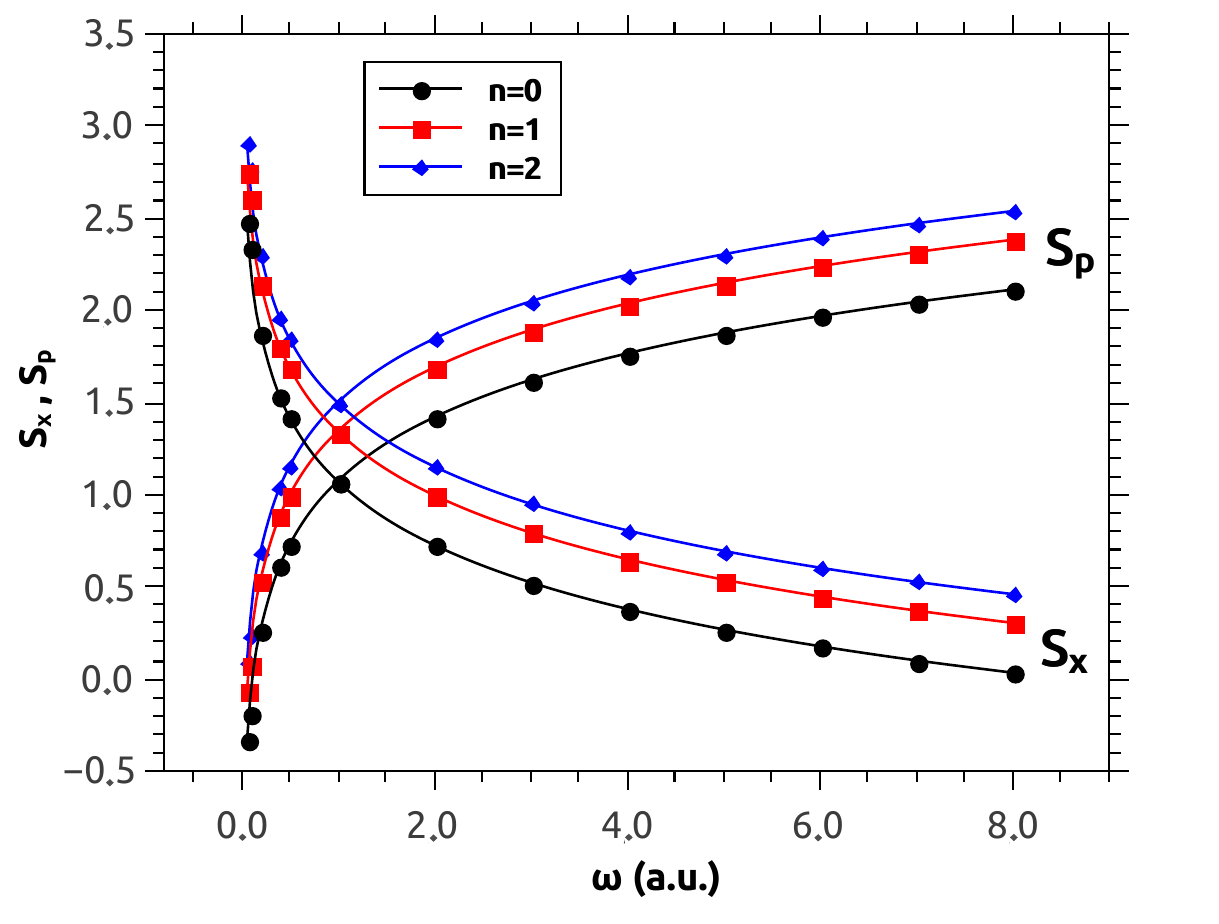}
\caption{Modified information entropies $S_x$ e $S_p$ as a function of $\omega$ for the harmonic oscillator to three lowest energy states.}
\label{comparacao_entropias_oscilador_g} % este é o nome da figura e sempre que você se referir a esta figura no texto digite apenas \ref{debian} que ela será referida corretamente.
\end{figure}

In figure~\ref{comparacao_entropias_oscilador_g}, the curves of $S_x$ e $S_p$ versus $\omega$ are presented. The crossing points of the curves for the three lowest energy states occur in $\omega ~=~1.000$ a.u. to $S_x = S_p = 1.0724$ for $n=0$, to $S_x~=~S_p~=~1.3427$ for $n=1$ and to $S_x~=~S_p~=~1.4986$ for $n=2$. For this $\omega$ value, the Hamiltonian is given by $H = \frac{1}{2} p^2+\frac{1}{2}x^2$ in a.u.. So the probability densities on position and momentum spaces have the same prevalence on the system, that is, are equally balanced. Furthermore, the values of $S_x$ and $S_p$ in crossing points increase with $n$ increment.

 % termina o ambiente figura

A analysis on probability densities is related to wave function in ground state, in this case a gaussian wave function. The variation of $\omega$ in the function modifies its amplitude, transforming the spreading of this probability distribution. Thus, we can consider the localization or delocalization of a particle.

\begin{figure}[h]
\centering
\includegraphics[width=16cm, height=6cm]{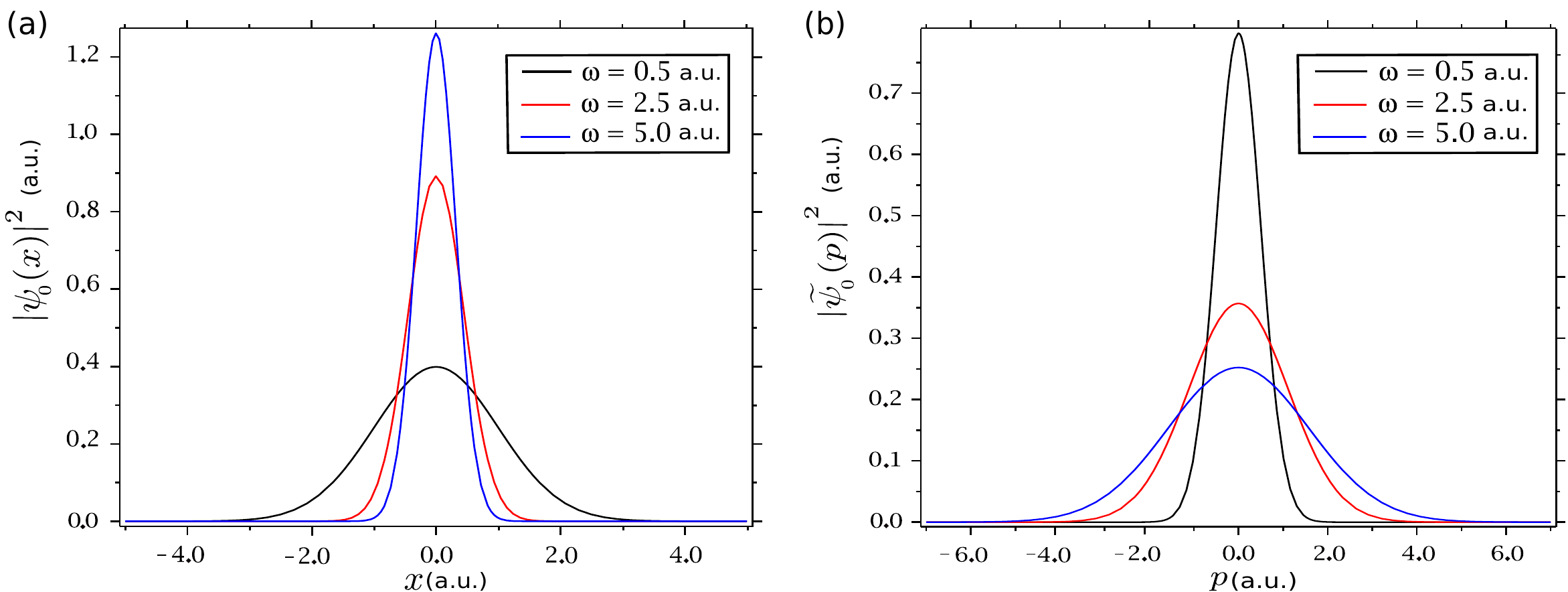}
\caption{For the harmonic oscillator (a) the probability densities on position space ${\arrowvert \psi_{0} (x)\arrowvert}^2$ and (b) the probability densities on momentum space ${\arrowvert \widetilde{\psi}_0(p)\arrowvert}^2$. The angular frequencies of $\omega$ are 0.5000~a.u., 2.5000~a.u. and 5.0000~a.u..}
\label{densidades_de_probabilidade_fundamental} 
\end{figure}

In figure~\ref{densidades_de_probabilidade_fundamental} we present the curves of ${\arrowvert \psi_{0} (x)\arrowvert}^2$ and ${\arrowvert \widetilde{\psi}_0(p)\arrowvert}^2 $ for distinct values of $\omega$. Decreasing values of $\omega$, the spreading of ${\arrowvert \psi_{0} (x)\arrowvert}^2$ increases, implying a growth of delocalization, \textit{i.e.}, $S_x$ value increases. The decrease of $\omega$ value comes together with a decrease of ${\arrowvert \widetilde{\psi}_{0} (p)\arrowvert}^2$ spreading, unveiling the uncertainty decreases on particle momentum, corresponding to a plunge in $S_p$ curves.

For $\omega$ values equal to 0.5000~a.u., 2.5000~a.u. and 5.0000~a.u. to the ground state, the $\Delta \hat{X}$ values are 1.0000 a.u., 0.4472 a.u. and 0.3162 a.u., and the $\Delta \hat{P}$ values are 0.5000 a.u., 1.1180 a.u. and 1.5811 a.u., respectively. Still, the $S_x$ values for such frequencies are 1.4189, 0.6142 and 0.2676, while the $S_p$ values are 0.7258, 1.5305 and 1.8771,  respectively. In all cases, $\Delta \hat{X} \Delta \hat{P}$ = 1/2 a.u. and $S_t=2.1447$. The standard deviation and information entropies are equivalent measures for Gaussian-type probability distribution, having the same qualitative behavior, satisfactorily describing the spread.

\begin{figure}[h]
\centering
\includegraphics[width=16cm, height=6cm]{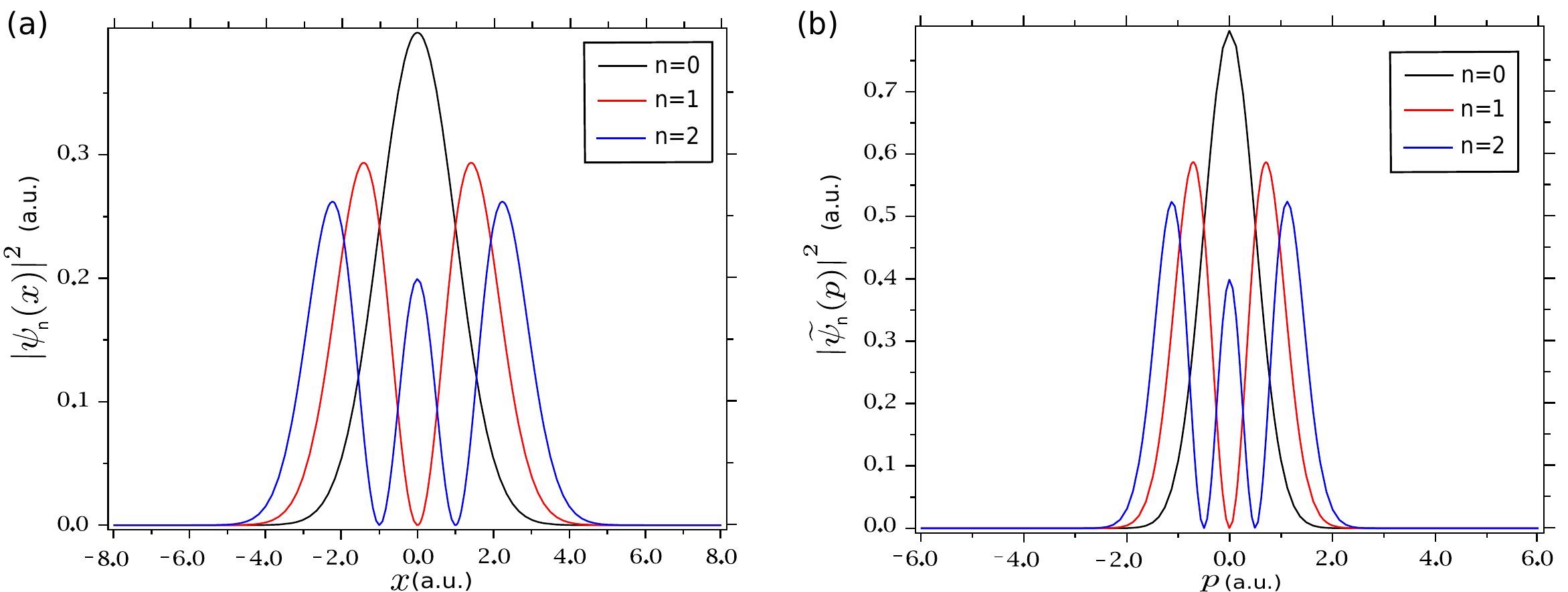}
\caption{For the harmonic oscillator the probability densities (a) in position space ${\arrowvert \psi_{0} (x)\arrowvert}^2$, ${\arrowvert \psi_{1} (x)\arrowvert}^2$ and ${\arrowvert \psi_{2} (x)\arrowvert}^2$ and the probability densities (b) in momentum space ${\arrowvert \widetilde{\psi}_0(p)\arrowvert}^2 $, ${\arrowvert \widetilde{\psi}_1(p)\arrowvert}^2 $ and ${\arrowvert \widetilde{\psi}_2(p)\arrowvert}^2 $. The angular frequency is $\omega$~=~0.5000 a.u..}
\label{densidades_de_probabilidade}
\end{figure}

Figure~\ref{densidades_de_probabilidade} presents the probability densities on position space ${\arrowvert \psi_{1} (x)\arrowvert}^2$, ${\arrowvert \psi_{2} (x)\arrowvert}^2$ and ${\arrowvert \psi_{3} (x)\arrowvert}^2$ and on momentum space ${\arrowvert \widetilde{\psi}_1(p)\arrowvert}^2 $, ${\arrowvert \widetilde{\psi}_2(p)\arrowvert}^2 $ and ${\arrowvert \widetilde{\psi}_3(p)\arrowvert}^2 $, for $\omega$~=~0.5000~a.u.. Note that the probability densities in position and momentum spaces increase their spreading in respective $x$ and $p$ domains of wave function with the increase of the quantum number. This represents an increasing uncertainty in position and momentum of a particle, a result that corroborates with increase of $S_x$ and $S_p$  with $n$ that is presented in table~\ref{energias_entropias_frequencias}.

\begin{figure}[b] % aqui começa o ambiente figura
\centering % este comando é para centralizar a figura
\includegraphics[width=9cm, height=7cm]{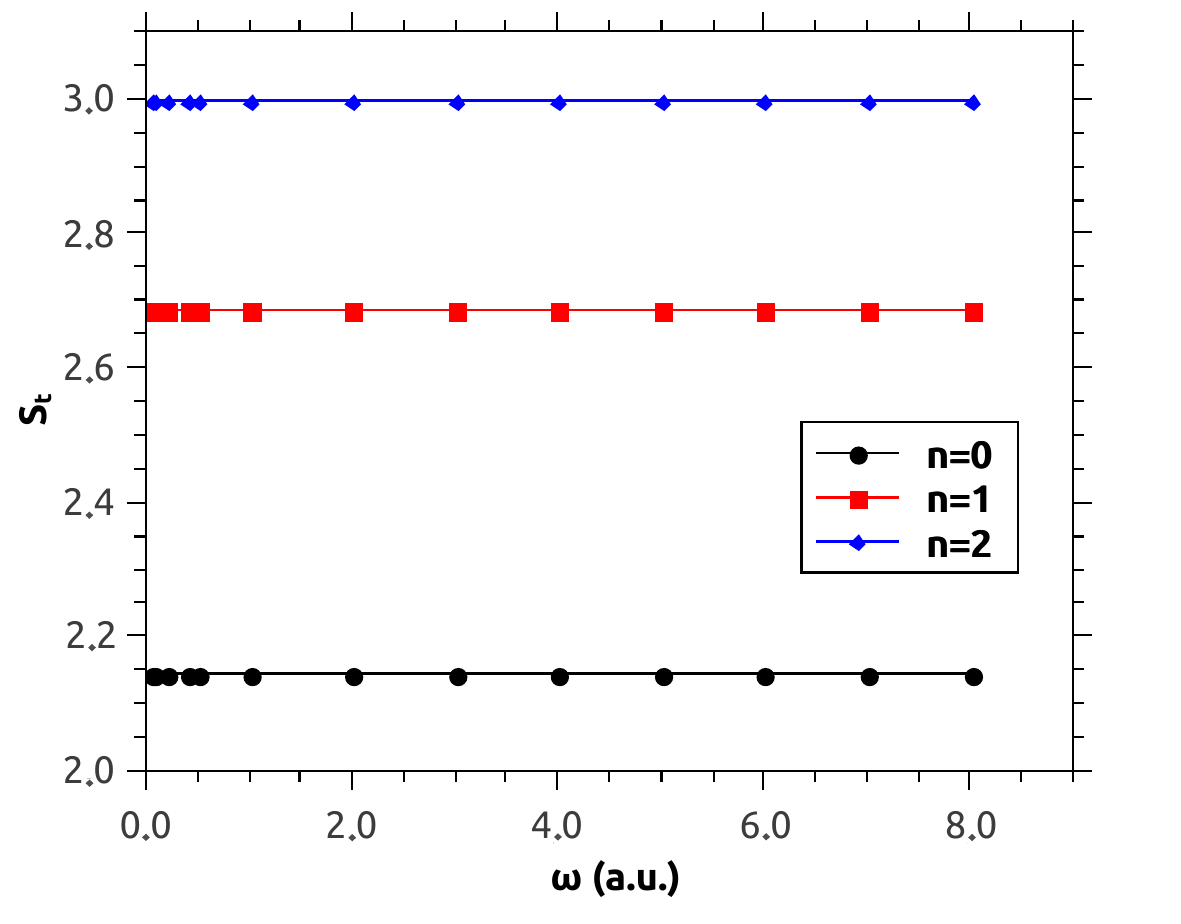}
\caption{Entropy sum $S_t$ as a function of $\omega$ for the harmonic oscillator to three lowest energy states.}
\label{comparacao_soma_entropica_oscilador_g} % este é o nome da figura e sempre que você se referir a esta figura no texto digite apenas \ref{debian} que ela será referida corretamente.
\end{figure} % termina o ambiente figura

The results obtained for the entropy sum as a function of $\omega$ are also arranged in table~\ref{energias_entropias_frequencias} and the behavior presented in figure~\ref{comparacao_soma_entropica_oscilador_g}. The $S_t$ value increases with $n$ increment. Moreover, despite of $S_x$ and $S_p$ changes, entropy sum maintains constant for each state of the system. Properties of Fourier transform, topic studied in undergraduate physics courses, can explain these results. Since Parseval's theorem states that Fourier transform is unitary, \textit{i.e.}, the integral of $\rho(x)$ has the same value as the integral of $\gamma(p)$, so the normalization constant of $\widetilde{\psi}(p)$ is equal to normalization constant of $\psi(x)$. The normalization constant of wave functions on position and momentum spaces depends on $\beta^{1/4}$. As $\psi(x)\equiv\psi(\sqrt{\beta}x)\propto \beta^{1/4}\phi(\sqrt{\beta}x)$, the scale property of Fourier transform states that $\widetilde{\psi}(p)\equiv \widetilde{\psi}(p/\sqrt{\beta})/\sqrt{\beta} \propto \widetilde{\phi}(p/\sqrt{\beta})/\beta^{1/4}$, where $\phi(x)$ and $\widetilde{\phi}(p)$ 
are the non-normalized wave functions on position and momentum spaces, respectively. So 
\begin{equation}
S_t=-\int_{-\infty}^{\infty}\int_{-\infty}^{\infty} dx \ dp \ \bar{\rho}(\sqrt{\beta}x)\ \bar{\gamma}(p/\sqrt{\beta}) \ \ln\left(\bar{\rho}(\sqrt{\beta}x)\ \bar{\gamma}(p/\sqrt{\beta})
\right). 
\end{equation}
Here, $\bar{\rho}(\sqrt{\beta}x)\propto \sqrt{\beta}|\phi(\sqrt{\beta}x)|^2$ and $\bar{\gamma}(p/\sqrt{\beta})\propto |\widetilde{\phi}(p/\sqrt{\beta})|^2/\sqrt{\beta}$. Replacing $\sqrt{\beta}x$ by $x$ and $p/\sqrt{\beta}$ by 
$p$, the entropy sum is written as
\begin{equation}
S_t=-\int_{-\infty}^{\infty}\int_{-\infty}^{\infty} dx \ dp \ \bar{\rho}(x) \ \bar{\gamma}(p) \ \ln\left(\bar{\rho}(x)\ \bar{\gamma}(p)
\right). 
\end{equation}
In this way, entropy sum is independent of $\beta$, and consequently of $\omega$ too.

%\clearpage 

\subsection{Infinite potential well}{\label{Infinite potential well result}}

\begin{figure}[b] % aqui começa o ambiente figura
\centering % este comando é para centralizar a figura
\includegraphics[width=9cm, height=7cm]{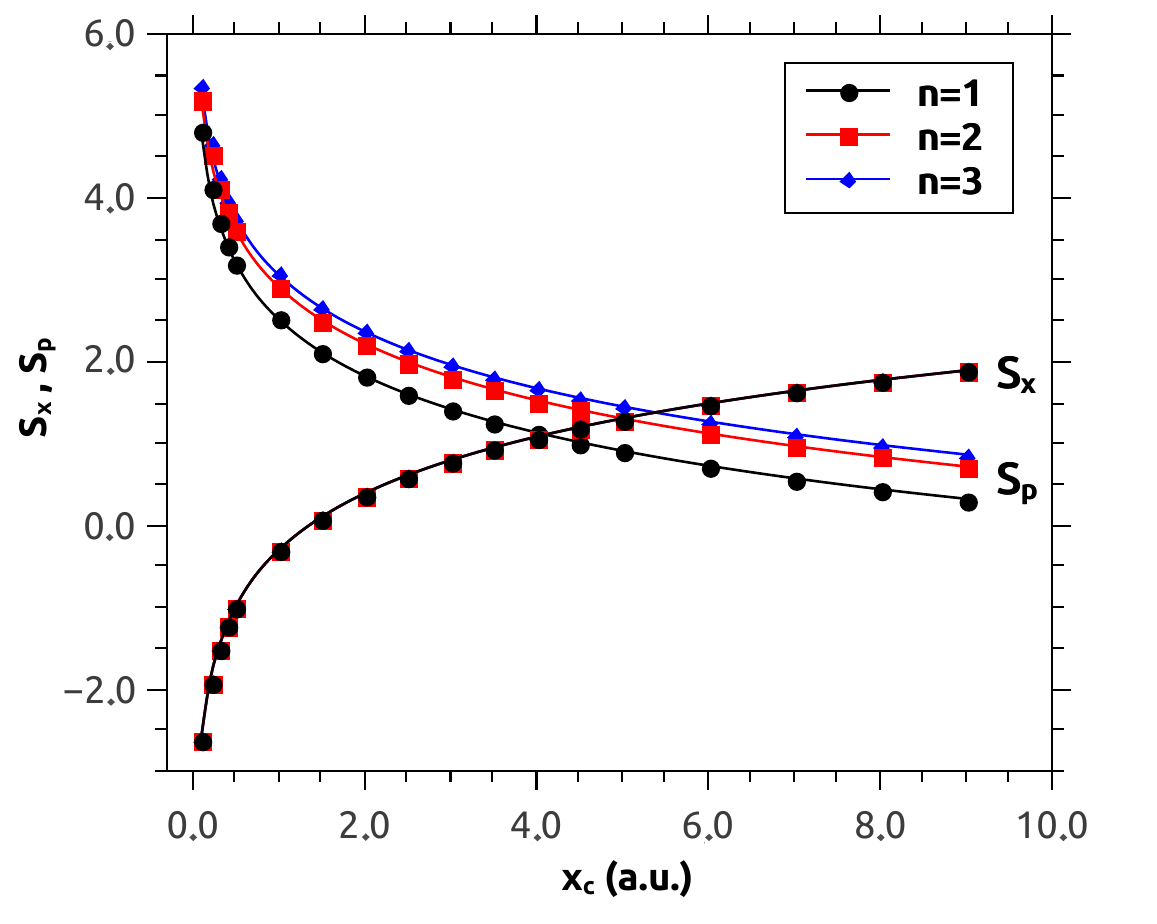}
\caption{Modified information entropies $S_x$ and $S_p$ for the confined particle in a box as a function of $x_c$ to three lowest energy states.}
\label{comparacao_entropias_g} % este é o nome da figura e sempre que você se referir a esta figura no texto digite apenas \ref{debian} que ela será referida corretamente.
\end{figure}

We investigated the particle confined in a box in ground ($n$~=~1) and also in two first excited ($n$~=~2~and~3) states by equations~(\ref{solucao_cos})~and~(\ref{solucao_sen}). The values of modified information entropies $S_x$ and $S_p$ for some values of $x_c$ are given in table~\ref{entropias_caixa_tabela} and displayed in figure~\ref{comparacao_entropias_g}. Note that the values of $S_x$ are identical for different quantum states. 

%\begin{landscape}
\begin{table}[!ht]
%\rowcolors{1}{}{lightgray}
%\footnotesize
\caption{Modified information entropies $S_x$ and $S_p$, besides of the entropy sum $S_t$ as a function of $x_c$ for the confined particle in a box to three lowest energy states. All values in atomic units system.} 
\label{entropias_caixa_tabela}
\centering
\begin{tabular}{ccccccccccccccccc}
\\ \hline \hline 
 & \vline &  & $S_x$	& & \vline	&		&$S_p$		&	& \vline	&	&$S_t$	&
\\ \hline
$r_c $ & \vline & $n$~=~1 & $n$~=~2	& $n$~=~3& \vline	&$n$~=~1		&$n$~=~2		& $n$~=~3	& \vline	&$n$~=~1	&$n$~=~2	&$n$~=~3
\\ \hline
0.1000& \vline &	-2.6094	&-2.6094	&-2.6094& \vline &	4.8215	&5.2164	&5.3625& \vline &	2.2120	&2.6070	&2.7531
\\ \hline
0.2000& \vline &	-1.9163	&-1.9163	&-1.9163& \vline &	4.1283	&4.5232	&4.6694& \vline &	2.2120	&2.6070	&2.7531
\\ \hline
0.3000& \vline &	-1.5108	&-1.5108 &-1.5108& \vline &	3.7229	&4.1178	&4.2639& \vline &	2.2120	&2.6070	&2.7531
\\ \hline
0.4000& \vline &	-1.2231	&-1.2231	&-1.2231& \vline &	3.4352	&3.8301	&3.9762& \vline &	2.2120	&2.6070	&2.7531
\\ \hline
0.5000& \vline &	-1.0000	&-1.0000	&-1.0000& \vline &	3.2120	&3.6070	&3.7531& \vline &	2.2120	&2.6070	&2.7531
\\ \hline
1.0000& \vline &	-0.3069	&-0.3069	&-0.3069& \vline &	2.5189	&2.9138	&3.0599& \vline &	2.2120	&2.6070	&2.7531
\\ \hline
1.5009& \vline &	0.0992	&0.0992	&0.0992	& \vline &2.1128	&2.5077 	&2.6538	& \vline &2.2120	&2.6070	&2.7531

\\ \hline
2.0000& \vline &	0.3863	&0.3863	&0.3863& \vline &	1.8257	&2.2207	&2.3668& \vline &	2.2120	&2.6070	&2.7531
\\ \hline
2.5000& \vline &	0.6094	&0.6094	&0.6094& \vline &	1.6026	&1.9975	&2.1437& \vline &	2.2120	&2.6070	&2.7531
\\ \hline
3.0000& \vline &	0.7918	&0.7918	&0.7918& \vline &	1.4203	&1.8152	&1.9613& \vline &	2.2120	&2.6070	&2.7531
\\ \hline
3.5000& \vline &	0.9459	&0.9459	&0.9459& \vline &	1.2661	&1.6611	&1.8072& \vline &	2.2120	&2.6070	&2.7531
\\ \hline
4.0000& \vline &	1.0794	&1.0794	&1.0794& \vline &	1.1326	&1.5275	&1.6737& \vline &	2.2120	&2.6070	&2.7531
\\ \hline
4.5000& \vline &	1.1972	&1.1972	&1.1972& \vline &	1.0148	&1.4098	&1.5559& \vline &	2.2120	&2.6070	&2.7531
\\ \hline
5.0000& \vline &	1.3026	&1.3026	&1.3026& \vline &	0.9094	&1.3044	&1.4505& \vline &	2.2120	&2.6070	&2.7531
\\ \hline
6.0000& \vline &	1.4849	&1.4849	&1.4849& \vline &	0.7271	&1.1221	&1.2682& \vline &	2.2120	&2.6070	&2.7531
\\ \hline
7.0000& \vline &	1.6391	&1.6391	&1.6391& \vline &	0.5730	&0.9679	&1.1140& \vline &	2.2120	&2.6070	&2.7531
\\ \hline
8.0000& \vline &	1.7726	&1.7726	&1.7726& \vline &	0.4394	&0.8344	&0.9805& \vline &	2.2120	&2.6070	&2.7531
\\ \hline
9.0050& \vline &	1.8909	&1.8909	&1.8909& \vline &	0.3211	&0.7160	&0.8621& \vline &	2.2120	&2.6070	&2.7531
\\ \hline \hline 
\end{tabular}
\end{table}
%\end{landscape}

 % termina o ambiente figura 

A mathematical explanation is given by normalization constants. Normalization constants for probability density $\rho(x)$ are the  same for different states, depends only on ${x_c}^{-1}$ for different wave functions in equations~(\ref{solucao_cos})~and~(\ref{solucao_sen}). In this way, the integrand of equation~(\ref{entropia_posicao_x}) for different states are related by a scaling factor equal $n$ to same confinement distance. For cosine solution (\ref{solucao_cos})
\begin{eqnarray}
 S_x=-\int_{-x_c}^{x_c}  dx \ |A|^2\cos^2(nx) \ln \left( {a_0} \ |A|^2\cos^2(nx) \right)=  \nonumber \\ \nonumber \\
 =-\frac{1}{n}\int_{-nx_c}^{nx_c}  dx \ |A|^2\cos^2(x) \ln  \left( {a_0} \ |A|^2\cos^2(x) \right) \ .
\end{eqnarray}
Here $nx$ was replaced by $x$. The integrand is a periodic function and it has full periods in interval $[-x_c,x_c]$. To integrate in new interval is the same as multiply the $S_x$ for fundamental state by $n$. However, the factor $1/n$ keeps the integral values for excited states the same as for fundamental one. For sine solutions~(\ref{solucao_sen}), an identical answer is obtained since sine and cosine functions are equal by a $\pi/2$ phase shift.

\begin{figure}[b]
\centering
\includegraphics[width=16cm, height=6cm]{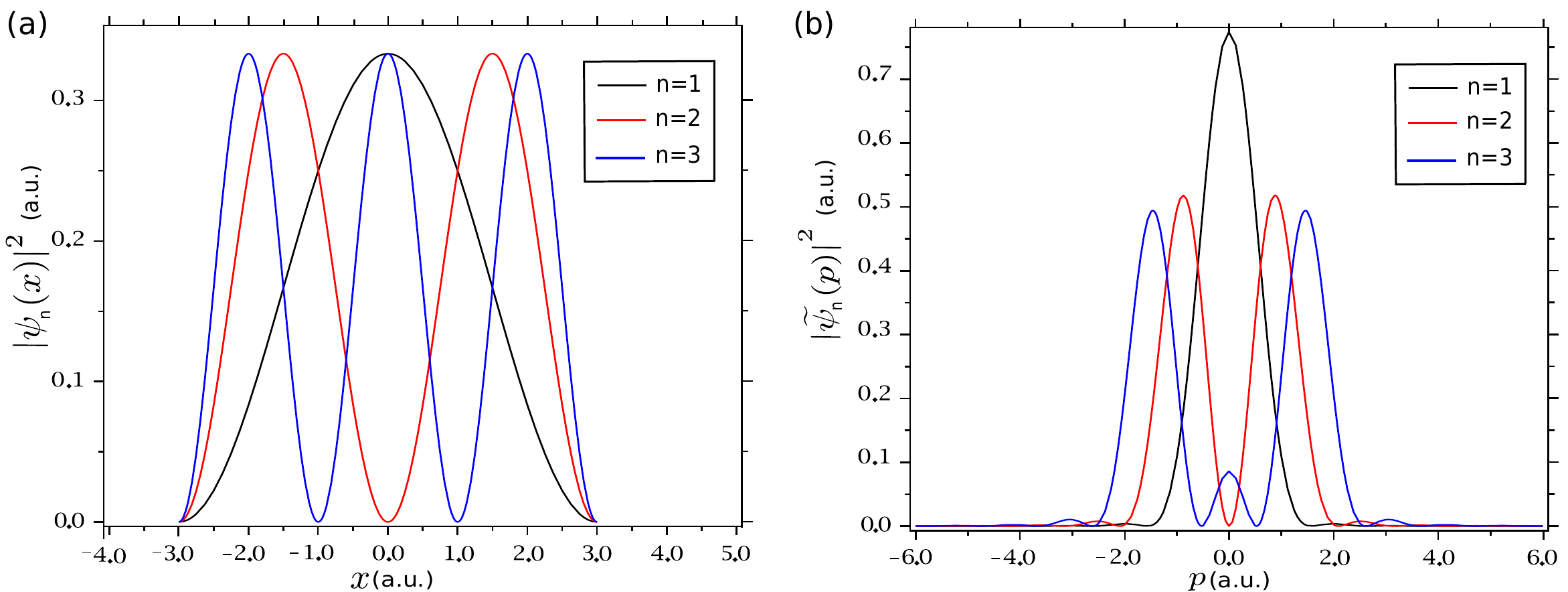}
\caption{For the confined particle in a box the probability densities (a) in position space ${\arrowvert \psi_{1} (x)\arrowvert}^2$, ${\arrowvert \psi_{2} (x)\arrowvert}^2$ and ${\arrowvert \psi_{3} (x)\arrowvert}^2$ and the probability densities (b) in momentum space ${\arrowvert \widetilde{\psi}_1(p)\arrowvert}^2 $, ${\arrowvert \widetilde{\psi}_2(p)\arrowvert}^2 $ and ${\arrowvert \widetilde{\psi}_3(p)\arrowvert}^2 $. The confinement distance is $x_c$~=~6.0000~u.a..}
\label{densidade_de_probabilidade_caixa}
\end{figure}

A qualitative explanation about behavior of modified entropies for different states can be obtained by analysis of the probability densities in position and momentum spaces. In figure~\ref{densidade_de_probabilidade_caixa} are presented the curves of these probability densities for the $x_c$~=~6.0000~a.u.. For example,  the curves ${\arrowvert \psi_{1} (x)\arrowvert}^2$, ${\arrowvert \psi_{2} (x)\arrowvert}^2$ and ${\arrowvert \psi_{3} (x)\arrowvert}^2$ are spread by same range of $x$ values (confinement limits $x=\pm x_c/2$), leading to equal values of $S_x$ for all states. On the other hand, the curves of ${\arrowvert\widetilde{\psi}_1(p)\arrowvert}^2 $, ${\arrowvert \widetilde{\psi}_2(p)\arrowvert}^2 $ and ${\arrowvert \widetilde{\psi}_3(p)\arrowvert}^2 $ are spread for increasing ranges of $p$ values increasing $n$, that explaining the increase in $S_p$ with $n$ increment. 

These results indicate that, in this sense, the information entropies $S_x$ and $S_p$ represent an uncertainty measure of the particle in the position and momentum spaces. The values of $S_x$ decrease when confinement becomes stronger, then the uncertainty in the particle's location decreases too. The values of $S_p$ increase with the confinement increment and affect more the states in an increasing order of energy. The behaviors of $S_x$ and $S_p$ are the same as found in reference~\cite{uncertainties_of_the_confined_harmonic}.

On the other hand, $\Delta \hat{X}$ values, calculated in $x_c$~=~6.0000~a.u. for $n=$ 1, 2 and 3 states, are 1.0845~a.u., 1.5950~a.u. and 1.6725~a.u., respectively. In contrast with the behavior of $S_x$, the $\Delta \hat{X}$ values vary with the $n$ increment. $\Delta \hat{X}$ is a measure of uncertainty that not only depends on $x_c$, but also consider a specific point of the probability distributions as reference, so variation with n is expected. On the other hand, $S_x$ is a entropic measure of uncertainty in the location of the particle in a box that only depends on $x_c$. 

In figure~\ref{comparacao_entropias_g} can be observed crossings between curves $S_x$ and $S_p$ for particular values of $x_c$. Approximately, for $n=1$ the crossing point occurs in $x_c=4.0000$ a.u., to $S_x=S_p=1.0794$. For $n=2$ and $n=3$ crossing points respectively occur in $x_c=5.0000$ a.u. to $S_x=S_p=1.3026$ and $x_c=5.3975$ a.u. to $S_x=S_p=1.3653$. This indicate that the values of the crossing points of the curves $S_x$ and $S_p$ increase with the increase of the quantum number.

\begin{figure}[!ht] % aqui começa o ambiente figura
\centering % este comando é para centralizar a figura
\includegraphics[width=9cm, height=7cm]{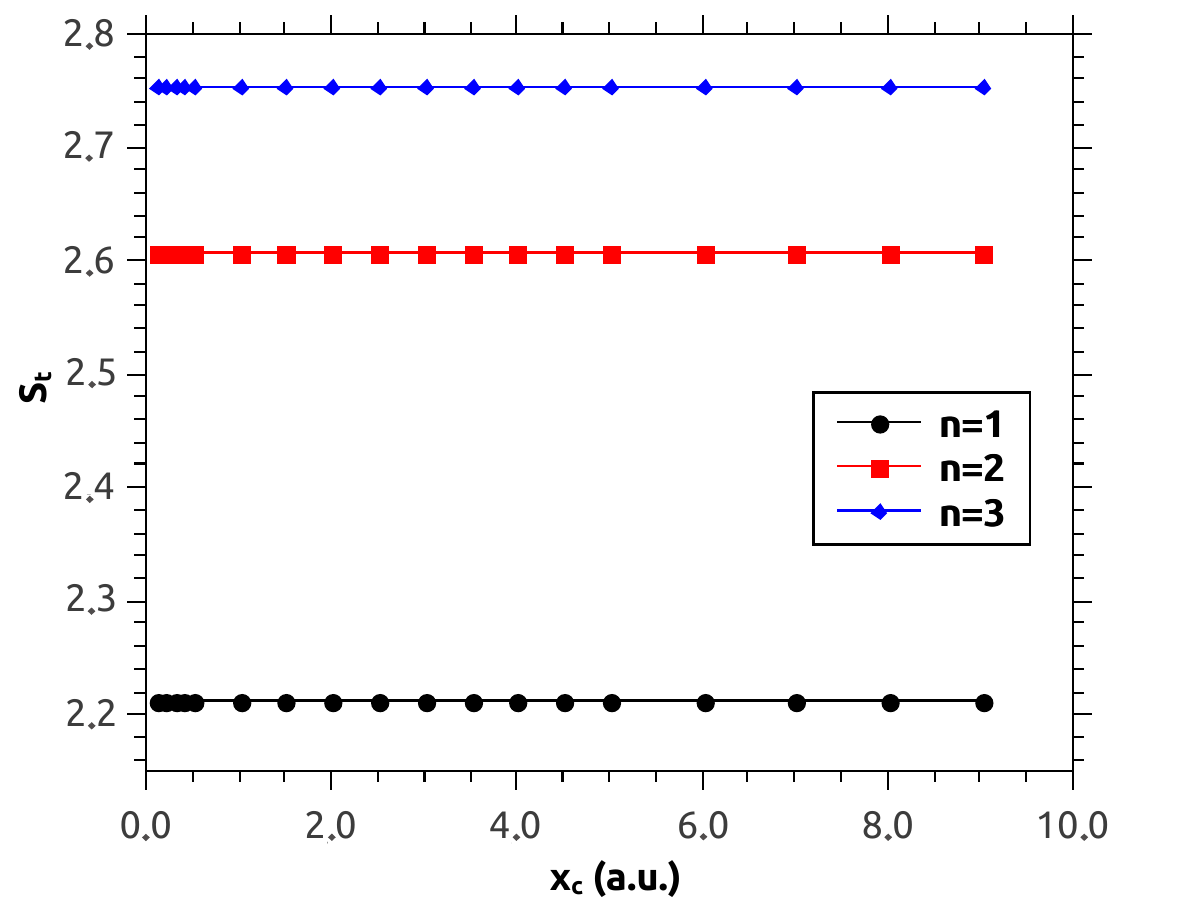} 
\caption{Entropy sum $S_t$ as a function of $x_c$ for the confined particle in a box to three lowest energy states.}
\label{comparacao_soma_entropica_g} % este é o nome da figura e sempre que você se referir a esta figura no texto digite apenas \ref{debian} que ela será referida corretamente.
\end{figure}

The values of the entropy sum $S_t$ as a function of $x_c$ are also found in table~\ref{entropias_caixa_tabela} and in 
figure~\ref{comparacao_soma_entropica_g} is presented its curve. For each quantum state, the value of $S_t$ remains constant despite of changes in $S_x$ and $S_p$. The reason is the same one we gave in harmonic oscillator case, but the parameter now is $x_c^{-1}$, in spite of explicitly $\beta^{1/2}$, or implicitly $\omega^{1/2}$. Additionally, $S_t$ assumes its least value for the ground state and it increases with $n$ increment; this is similar to results of reference~\cite{gardre1985}. The entropic uncertainty relation is respected for the different values of $x_c$ and $n$ since it is above its minimum defined in equation~(\ref{Stx}).

 % termina o ambiente figura 

Note in tables~\ref{energias_entropias_frequencias} and~\ref{entropias_caixa_tabela} that some of $S_p$ and $S_x$ values are negative. These values correspond to regions where the probability densities are highly localized. This fact can also be observed in the spherically confined hydrogen atom~\cite{benchmark_hidrogenio, wallas_fred_internacional} and systems with static screened Coulomb potential~\cite{information_static_screened_Coulomb_potential}.

\subsection{Considerations about both systems}

The probability densities of excited quantum states have a nodal structure (points in the space of positions that assume null values) that increase with increasing quantum number $n$. For the harmonic oscillator the $n$-th excited state has $n$ nodes, in figure~\ref{densidades_de_probabilidade} one can appreciate the nodal structure in (a) position and (b) momentum space for the first three quantum states. For the confined particle in a box the $n$-th excited state has $n-1$ nodes (the boundary conditions are not nodes), in figure~\ref{densidade_de_probabilidade_caixa} we observed the nodal structure in (a)~position and (b)~momentum space for $n=$ 1, 2 and 3 states.

Generally the increase in the number of nodes causes spreading of the probability distribution in position space. For the harmonic oscillator this causes an increase in the $S_x$ values with $n$ increments in table~\ref{energias_entropias_frequencias}. However, such spreading is avoided by the infinite potential barriers of the box in which the particle is confined, thus promoting values of $S_x$ independent of the quantum number in table~\ref{entropias_caixa_tabela}. This aspect again highlights the difference between the standard deviation and information entropy, because $\Delta \hat{X}$ increases for both considered systems. On the other hand, the $S_p$ values of the studied quantum systems always increase when $n$ also increases, and the entropic uncertainty relation is mantained valid.

Another interesting observation is about the influence of the nodal structure of probability densities on the value of $S_t$. In either cases, $S_t$ is always constant for each $n$ and their values increase when the quantum number $n$ grows. Moreover, for ground state the $S_t$ value of the confined particle in a box is of 2.2120, that is, greater than the harmonic oscillator which is 2.1447 (minimum value of the entropic uncertainty relation). For excited states, the $S_t$ values are smaller for the confined particle in a box than respective harmonic oscillator due to the invariance of $S_x$ values in relation to the $n$ increments (see tables~\ref{energias_entropias_frequencias}~and~\ref{entropias_caixa_tabela} for numerical values).

We can also note in both systems that the growth rate of  $S_t$ decreases when the quantum number $n$ increases. Meantime, the $\Delta \hat{X} \Delta \hat{P}$ quantity varies linearly with $n$ (or nodes number) for the harmonic oscillator and near linearly for the confined particle in a box.\footnote{In this case, $\Delta \hat{X} \Delta \hat{P}$ grows linearly with the quantum number when $n$ tends to infinity.} This feature makes the value of $\Delta \hat{X} \Delta \hat{P}$ smaller for the harmonic oscillator than  the confined particle in a box for the first three quantum states, when then an inversion occurs.

\section{Conclusions}{\label{conclusions}}

We pedagogically explored some elements of the information theory and provided a link with the quantum theory through of the information entropies. The physical significance of these entropies were discussed and a comparison with the standard deviations and the Kennard's relation was performed. 

In particular, we apply the modified entropic expressions in one-dimensional quantum systems to ground and two first excited states. For the $S_x$, $S_p$ and $S_t$, we provide an original explanation of their behaviors analyzing the probability densities and by means of the normalization constants and properties of Fourier transform.

In the harmonic oscillator problem, the value of $S_x$ increases and $S_p$ decreases when $\omega$ value reduces. An increase of $S_x$ and $S_p$ implies an increase in the uncertainty of position and momentum of a particle, \textit{i.e.}, an increase in the spreading of the respective probability densities in their domains. For the confined particle in a box, the values of $S_x$ are equal for all considered quantum states and decrease when confinement becomes stronger. The values of $S_p$ increase with the $n$ increment and advance of confinement.

For studied physical systems, despite of the changes in $S_x$ and $S_p$, the value of $S_t$ keeps constant for each quantum state and takes its smallest value for the ground state. The value of $S_t$ for the harmonic oscillator ground state is the minimum value of the entropic uncertainty relation. But, for the excited states the values of $S_t$ are smaller for the confined particle in a box. The entropic uncertainty relation is respected for the considered systems.

Moreover, for the ground state of the harmonic oscillator, the standard deviation and information entropies are equivalent measures of the spread of the probability distributions or the quantum uncertainty. On the other hand, for the confined particle in a box the $\Delta \hat{X}$ values, contrasting with the $S_x$ values, vary with $n$ increment. These results highlight the different characteristics of $\Delta \hat{X}$ ($\Delta \hat{P}$) and $S_x$ ($S_p$) as measures of spread or uncertainty. This fact already justifies the interest in introducing to undergraduate physics students the ideas of information entropy and theory of information in the quantum mechanics context.   

\ack

This work has been supported by the Brazilian agencies CAPES (Coordena\c{c}\~ao de Aperfei\c{c}oamento de Pessoal de N\'ivel Superior) and CNPq (Conselho Nacional de Desenvolvimento Cient\'ifico e Tecnol\'ogico) through grants to the authors. The authors thank the referees for careful reading of the manuscript and for helpful comments and suggestions.

\appendix
\setcounter{section}{1}

\section*{Appendix A}{\label{AppendixA}}

In the context of the quantum theory the dimensional adequate information entropies on position, $S_x$ , and on momentum, $S_p$, spaces for one-dimensional systems are~\cite{wallas_fred_internacional}: 
\begin{equation}
S_x =  - \int dx \ \rho(x)\ln( ({a_0}) \ \rho(x)) \ 
\label{entropia_posicao_completa}
\end{equation}
and
\begin{equation}
S_p =  - \int dp \ \gamma(p)\ln \left( {\left( \frac{\hbar}{a_0} \right)} \ \gamma(p)  \right) \ . 
\label{entropia_momento_completa}
\end{equation}

Using the Eqs.~(\ref{entropia_posicao_completa}) and (\ref{entropia_momento_completa}) to compose the entropy sum we have
\begin{equation}
S_t = S_x  + S_p = - \int  \rho(x)\ln( ({a_0}) \ \rho(x))dx - \int \gamma(p)\ln \left( {\left( \frac{\hbar}{a_0} \right)} \ \gamma(p)  \right) dp \ .
 \label{st1}
\end{equation}
Rewriting the expression to $S_t$ we get
\begin{equation}
S_t = -\int \int \ dx \ dp \  [\rho(x)  \ \gamma(p)] \left[ \ln( ({a_0})\rho(x) ) + \ln \left( {\left( \frac{\hbar}{a_0} \right)}\gamma(p) \right) \right] \ .
\end{equation}
Making a convenient choice and working with the properties of the logarithmic function we can write
\begin{equation}
S_t = -\int \int \ dx \ dp \  [\rho(x)  \ \gamma(p)] \left[ \ln\left( ({a_0})  \left( \frac{\hbar}{a_0} \right)   \rho(x)\gamma(p) \right) \right] \ .
\end{equation}
Thus, we find the expression of the entropy sum in terms of the fundamental constant $ \hbar$, that is,  
\begin{equation}
S_t = -\int \int \ dx \ dp \  [\rho(x)  \ \gamma(p)] \left[ \ln( \hbar \ \rho(x) \ \gamma(p) )\right] \ .
\label{St}
\end{equation}

From the entropy sum is derived the entropic uncertainty relation. The inequality of Eq.~(\ref{Stx}) is obtained by defining the (q, n)-norm of the Fourier transform. For a further study see the reference~\cite{relacaodeincertezainformacao}.

\appendix
\setcounter{section}{2}

\section*{Appendix B}{\label{Appendix}}

The modified information entropies $S_x$ and $S_p$ one dimension are defined by  equations~(\ref{entropia_posicao_x})~and~(\ref{entropia_momento_x}). A dimensional analysis of $S_x$, where $a_0$ and $dx$ has dimensions of length $[L]$, and $\rho(x)$ dimension of inverse of length $1/[L]$ we have 
\begin{equation}
S_x [=] [L] \left[  \frac{1}{L} \right]   \ln \left[ [ L] \left[  \frac{1}{L} \right]\right]  \ , 
\label{shannon_posicao_dimensao}
\end{equation}
consequently $S_x$ is a dimensionless quantity. Here $[=]$ refers to dimensional equality.

Realizing the dimensional analysis of $S_p$, where $\hbar/a_0$\footnote{A dimensional analysis of 
$\left( \frac{\hbar}{a_0} \right)$ we have: $\left[\frac{\hbar}{a_0}\right] [=] \left[\left( \frac{[E][T]}{[L]} \right)\right][=] 
\frac{[M][L]^2[T]}{[L][T]^2}[=]\frac{[M][L]}{[T]}[=][P]$.} and $dp$ has dimension of momentum $[P]$, and $\gamma(p)$ dimension of inverse 
of momentum $1/[P]$ is
\begin{equation}
S_p [=] [P] \left[  \frac{1}{P} \right]   \ln \left[ [ P] \left[  \frac{1}{P} \right]\right]  \ , 
\label{shannon_momento_dimensao}
\end{equation}
so, also $S_p$ is a dimensionless quantity. 

Insofar the entropy sum corresponds to a addition between two dimensionless quantities, it is also a dimensionless one. A dimensional analysis of $S_t$, where $[\hbar]=[E][T]$, being $[E]$ the dimension of energy and $[T]$ the dimension of time, we have
\begin{equation}
S_t [=][L][P] \left[\frac{1}{L}\right] \left[\frac{1}{P}\right] \ln \left[[E][T]  \left[\frac{1}{L}\right] \left[\frac{1}{P}\right]\right][=][1] \ .
\end{equation}
Remembering that  $[E]=\frac{[M][L]^2}{[T]^2}$, where $[M]$ is the dimension of mass, we have  
\begin{equation}
S_t [=][1] \ln \left[ \frac{[M][L]^2}{[T]^2}[T] \left[\frac{1}{L}\right] \left[\frac{1}{P}\right] \right] \ .
\end{equation}
And, $[P]=\frac{[M][L]}{[T]}$, \textit{i.e.},
\begin{equation}
S_t[=][1] \ln \left[ \frac{[M][L]^2}{[T]^2}[T] \left[\frac{1}{L} \right] \left[\frac{1}{\frac{[M][L]}{[T]}}\right] \right] \ .
\end{equation}
Finally,
\begin{equation}
S_t[=][1] \ln[1] \ .
\label{dis1}
\end{equation}

Using atomic units in the modified relationships recover the conventional way, but, now with a convenient dimensional discussion.

\section*{References}

\bibliography{references}
\bibliographystyle{unsrt}

\end{document}